\documentclass[aps,prd,twocolumn,superscriptaddress,amsmath,amssymb,floatfix,nofootinbib]{revtex4}
\usepackage{xcolor,graphicx,verbatim,bm}
\usepackage[colorlinks=true,citecolor=blue,filecolor=blue,linkcolor=blue,urlcolor=blue,pdftex]{hyperref}
\usepackage[utf8]{inputenc}

\begin{document}

\title{Imprints of neutrino-pair flavor conversions on nucleosynthesis in ejecta from neutron-star merger remnants}

\author{Meng-Ru Wu}
 \email{mwu@gate.sinica.edu.tw}
 \affiliation{Niels Bohr International Academy, Niels Bohr Institute, Blegdamsvej 17, 2100 Copenhagen, Denmark}
 \affiliation{Institute of Physics, Academia Sinica, Taipei, 11529, Taiwan}
 \affiliation{Institute of Astronomy and Astrophysics, Academia Sinica, Taipei, 10617, Taiwan}
\author{Irene Tamborra}
 \email{tamborra@nbi.ku.dk}
 \affiliation{Niels Bohr International Academy, Niels Bohr Institute, Blegdamsvej 17, 2100 Copenhagen, Denmark}
 \affiliation{DARK, Niels Bohr Institute,  Juliane Maries Vej 30, 2100, Copenhagen, Denmark}
\author{Oliver Just}
\email{oliver.just@riken.jp}
\affiliation{Astrophysical Big Bang Laboratory, RIKEN, Saitama 351-0198, Japan} 
\affiliation{Max-Planck-Institut f\"ur Astrophysik, Karl-Schwarzschild-Str.~1, 85748 Garching, Germany}
\author{Hans-Thomas Janka}
 \email{thj@mpa-garching.mpg.de}
 \affiliation{Max-Planck-Institut f\"ur Astrophysik, Karl-Schwarzschild-Str.~1, 85748 Garching, Germany}

\date{\today}

\begin{abstract}
The remnant of neutron star mergers is dense in neutrinos.  
By employing inputs from one hydrodynamical simulation
of a binary neutron star merger remnant with a black hole 
of $3\ M_\odot$ in the center, dimensionless spin
parameter $0.8$ and an accretion torus of $0.3\ M_\odot$, 
the neutrino emission properties are investigated as the merger remnant evolves. 
Initially, the  local number 
density of $\bar{\nu}_e$ is larger than that of $\nu_e$ everywhere above the remnant.
Then, as the torus approaches 
self-regulated  equilibrium, the local abundance of neutrinos overcomes 
that of antineutrinos in a funnel around the polar region. 
The region where the fast pairwise flavor conversions
can occur shrinks accordingly as time evolves. Still, we find that 
fast flavor conversions do affect most of the neutrino-driven ejecta.
Assuming that fast flavor
conversions lead to flavor equilibration, 
a significant enhancement of 
nuclei with mass numbers $A>130$ is found  as well as 
a change of the lanthanide mass fraction
by more than a factor of a thousand. Our findings  hint towards a 
 potentially relevant role of neutrino flavor oscillations 
for the prediction of the kilonova (macronova)
light curves and motivate further work in this direction.

\end{abstract}


\maketitle

\section{Introduction}
Compact binary mergers originate from the coalescence of a neutron star (NS) 
with another NS or a black hole (BH). 
They have long been considered to be precursors of  
short gamma-ray bursts (sGRBs), main hosts of the nucleosynthesis of heavy 
elements~\cite{Eichler:1989ve,1974ApJ...192L.145L},
and sources of gravitational waves (see \cite{Baiotti:2016qnr} for a recent review 
and references therein). 
The radioactive decay of the synthesized 
neutron-rich nuclei has also been assumed 
to power electromagnetic transients, 
called kilonovae or macronovae~\cite{Li:1998bw,Kulkarni:2005jw,Metzger:2010sy} 
(see also, e.g., Refs.~\cite{Metzger:2016pju,Metzger:2017wot} for recent reviews). 
These conjectures have recently been  confirmed by the 
detection of the GW170817 event and its related electromagnetic 
counterparts~\cite{Monitor:2017mdv,GBM:2017lvd,TheLIGOScientific:2017qsa} 
(see also Ref.~\cite{Metzger:2017wot} and references therein for the kilonova interpretation).
Beyond the kilonova associated to the GW170817 event, a few more potential 
kilonova candidates have been identified
through the infrared excess linked to the sGRB afterglows~\cite{Berger:2013wna,Tanvir:2013pia,Jin:2015txa,Jin:2016pnm}. 
The upcoming increasing statistics in the detection of such events, 
both for what concerns gravitational waves  and  electromagnetic counterparts, will  
greatly improve our understanding of compact binary mergers
and provide rich implications on their physics~\cite{Bloom:2009vx,Fernandez:2015use}. 
 
In binary NS mergers or NS-BH mergers, neutrinos can be 
copiously generated due to the violent collision of the two NSs and
the presence of the hot and dense post-merger massive NS or BH accretion disk. 
Similarly to core-collapse supernovae (CCSNe)~\cite{Mirizzi:2015eza},
neutrinos play an important role in mergers as they 
dominate the cooling of the merger remnants, change the composition of the ejecta, 
and affect the nucleosynthesis outcome in the ejecta and eventually 
the electromagnetic light curves~\cite{Ruffert:1996by,Foucart:2015vpa,Perego:2014fma,Wanajo:2014wha, 
Richers:2015lma,Just:2014fka}.
Neutrinos can also contribute to energizing 
sGRBs via pair annihilation above the BH accretion 
disk~\cite{Eichler:1989ve,Narayan:1992iy,Berger:2013jza,Fong:2013lba,Just:2015dba}. 

The exploration of the role of neutrinos 
in the merger remnants is still preliminary due to the highly demanding 
computational requirements for fully three-dimensional, general-relativistic 
magnetohydrodynamical modeling with detailed neutrino transport. On the other hand, 
various simulations show a generic feature which is the protonization 
of the merger remnant 
(i.e., more $\bar\nu_e$'s  than $\nu_e$'s are emitted).
The protonization of the merger remnant 
has peculiar implications for the flavor 
conversion of neutrinos.
For example, the so-called ``matter-neutrino resonance'' 
(MNR)~\cite{Malkus:2014iqa,Malkus:2012ts,Wu:2015fga,Zhu:2016mwa,Frensel:2016fge,Tian:2017xbr} 
is expected to occur. The MNR is due to the cancellation 
of the matter potential describing the interactions of neutrinos with electrons 
($\nu$-$e$) and the neutrino-neutrino potential ($\nu$-$\nu$). 
The MNR is typical of  merger remnants;
it does not occur in, e.g.~CCSNe, unless physics
beyond the Standard Model 
is involved~\cite{Wu:2015glr,Stapleford:2016jgz}.

Given the nature of the $\nu$-$\nu$ interaction, the neutrino angular distribution 
may affect the overall flavor conversion. In particular, ``fast'' pairwise flavor 
conversions~\cite{Sawyer:2005jk,Sawyer:2008zs,Sawyer:2015dsa} strongly 
depend on the angular distribution of the electron neutrino lepton 
number (ELN)~\cite{Dasgupta:2016dbv,Izaguirre:2016gsx,Wu:2017qpc} and 
can develop 
on time scales of  $(G_F |n_{\nu_e}-n_{\bar{\nu}_e}|)^{-1} \simeq 
\mathcal{O}(10)$~cm at the vicinity of the neutrinosphere, 
with $G_F$ being the Fermi constant and $n_{\nu_e}$
($n_{\bar{\nu}_e}$) the local $\nu_e$ ($\bar{\nu}_e$) number density. 
For any location above the neutrino emitting surface at a given time, 
fast conversions may occur if the ELN angular distribution 
has crossings, i.e., the ELN angular distribution is
positive in a certain solid-angle range, but negative in another one. 
 As the temporal and spatial scales during which fast conversions develop
are small compared to the size of the astrophysical object,
they may lead to flavor equilibration. 

The role of the angular distribution of neutrinos in the context of  flavor conversions
in merger remnants has been investigated in Ref.~\cite{Wu:2017qpc} for the first time. 
A simple two-disk neutrino emission model, with the decoupling region of $\bar{\nu}_e$ 
sitting within that of $\nu_e$, has been used. This model, however, 
was stationary and assumed uniform properties for neutrinos everywhere on 
the neutrino emitting surfaces. 
Moreover, the geometrical shape of the neutrino emitting surfaces 
was approximated with flat disks.  
Due to the remnant protonization and the emission geometry, 
the authors of Ref.~\cite{Wu:2017qpc} found that favorable conditions for flavor instabilities exist 
 for any point above the $\nu_e$ emission surfaces.

\begin{figure}[]
  \includegraphics[angle=0,width=1.01\columnwidth]{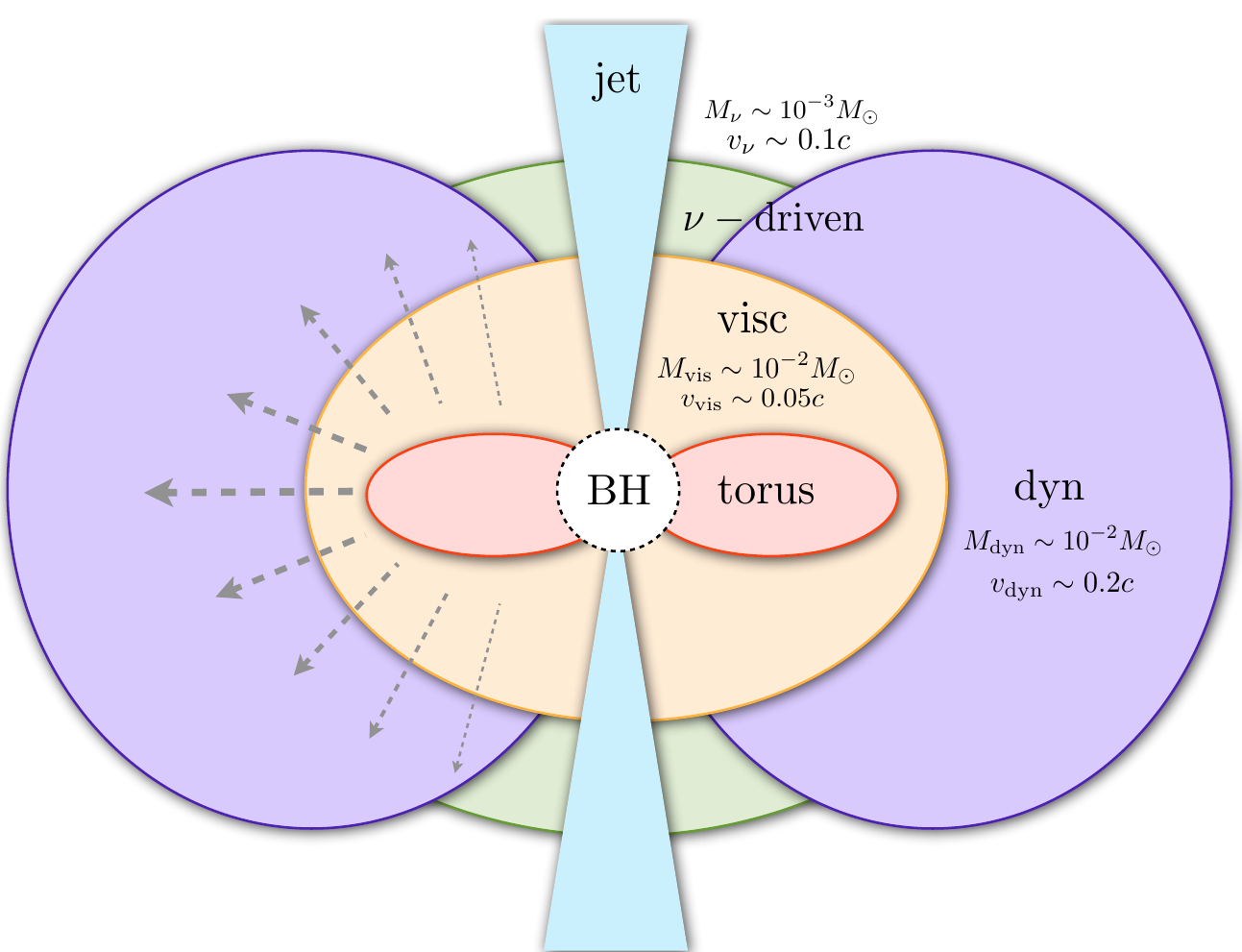}     
  \caption{Schematic representation of a BH-torus remnant and its ejecta resulting from 
    a NS-NS or a NS-BH merger. The dynamical ejecta (``dyn'', violet-shaded area) is the earliest matter
    outflow, followed by the neutrino-driven component (``$\nu$-driven'', green-shaded area) and finally the 
    viscously driven ejecta (``visc'', orange-shaded area).  For each of the aforementioned
    outflow components, characteristic values are reported for the ejecta mass and
    velocity. Neutrino-driven winds may dominate the ejecta in a cone centered at the polar axis
    with half-opening angle of about $10$--$40^o$. The picture sketched here for a BH-torus remnant is
    qualitatively similar to the case of a NS-torus system that may form after a NS-NS merger, except
    that more massive neutrino-driven winds are expected with a central NS. }
  \label{fig:toymodel}
\end{figure}

In this work, we intend to further explore the role of fast conversions in merger 
remnants and investigate their potential impact on the synthesis of elements by
adopting a  realistic remnant configuration based on numerical simulations from Ref.~\cite{Just:2014fka}. 
Figure~\ref{fig:toymodel} sketches the typical geometry 
of the merger remnant and the different components of the matter 
ejected during and after the merger.
The central compact object can be a BH or, in the case of a binary NS merger,
a NS that may collapse to a BH at some later time.
The outermost layer consists of the earliest dynamical ejecta which become
unbound via tidal torques during the merger or through the 
violent collision of two NSs from the contact interface within $\lesssim 10$~ms
after the merging.
The total amount of these dynamical ejecta can be up to 
$10^{-3}$--$10^{-1}$~$M_\odot$~\cite{Hotokezaka:2012ze,Bauswein:2013yna,Sekiguchi:2015dma,Radice:2016dwd,Rosswog:2016dhy,Foucart:2014nda}.
Neutrinos have a negligible role for the ejecta in the equatorial plane
but may greatly influence the polar ejecta~\cite{Sekiguchi:2015dma,Foucart:2016rxm}.

Following the merger, a remnant accretion disk of up to $\sim 10^{-1}$~$M_\odot$
surrounding the central massive NS or BH can form.
Recent hydrodynamical simulations based on idealized initial conditions show that
$\gtrsim 20$\% of the initial disk mass can be further ejected
via various mechanisms.
In the first few hundred milliseconds, neutrinos coming from the 
hot and dense region of the inner disk and/or the central 
massive NS  (prior to its collapse to a BH) can cause a neutrino-driven wind, 
dominantly around the polar region with a total mass of 
$\sim 10^{-3}$~$M_\odot$~\cite{Perego:2014fma,Just:2014fka}.
On a longer time scale of a few seconds, $\sim 10^{-2}$~$M_\odot$ 
can be further ejected by viscous heating and 
nuclear recombination~\cite{Fernandez:2013tya,Metzger:2014ila,Just:2014fka,Wu:2016pnw,Lippuner:2017bfm,Siegel:2017nub}. 
Both components are shown schematically in Fig.~\ref{fig:toymodel} in green and orange respectively.
One can see that the neutrino-driven ejecta may become the dominant component 
in the polar direction.
Consequently, the potential change of the relative abundance of neutrinos 
of different flavors due to fast flavor conversions may eventually affect 
the nucleosynthesis outcome in that region.

Throughout the rest of the paper, we focus on the neutrino-driven ejecta
in the post-merger phase.
For the first time, we study the evolution of the neutrino emission properties
by adopting inputs from one hydrodynamical simulation 
of a post-merger BH accretion disk. 
Based on model M3A8m3a5 of Ref.~\cite{Just:2014fka},
we perform a flavor stability analysis for several time snapshots to pinpoint the eventual
occurrence of fast pairwise conversions. 
In order to gauge the importance 
of neutrino flavor conversions in the remnant, we then quantify whether 
the flavor equipartition induced by fast pairwise conversions may be 
responsible for a non-negligible effect on the synthesis of heavy elements 
in the neutrino-driven ejecta. 
 
The manuscript is organized as follows. In Sec.~\ref{sec:nuproperties}, we 
first study and characterize the neutrino emission properties obtained in the hydrodynamical 
simulation of  model M3A8m3a5 presented in 
Ref.~\cite{Just:2014fka}. In Sec.~\ref{sec:nuoscillations}, we 
perform a time-dependent stability analysis. In Sec.~\ref{sec:nucleosynthesis} we 
discuss how the possible occurrence of flavor equipartition may affect the 
nucleosynthesis in the neutrino-driven ejecta from the merger remnants. 
Our conclusions and an outlook are presented in Sec~\ref{sec:conclusions}. 
Further details on the flavor stability analysis
are reported in Appendix~\ref{sec:app-flavor}.

\section{Neutrino emission properties}\label{sec:nuproperties}
In this section, we describe the main features of the BH-torus 
evolution of  model M3A8m3a5  with a central BH of $3\ M_\odot$, 
dimensionless BH spin parameter $0.8$ and torus of $0.3\ M_\odot$. 
Note that we have chosen  model M3A8m3a5 from the ones presented 
in Ref.~\cite{Just:2014fka}, as this is the one with the largest fraction of neutrino-driven
ejecta due to its large torus mass. 
Therefore, this case represents an optimistic scenario
to explore the role of neutrino flavor conversions in the merger remnant.
Special attention is dedicated to the evolution of the ELN and its
angular distribution, 
the crucial quantity in the context of fast flavor conversions.

\subsection{Evolution  of binary neutron star merger remnants}
The authors of Ref.~\cite{Just:2014fka} adopted a pseudo-Newtonian
Artemova-Novikov gravitational potential~\cite{Artemova:1997ra}
and energy-dependent neutrino transport scheme coupled to the 
Navier-Stokes equations with a Shakura-Sunyaev viscosity prescription to 
model the post-merger long-time evolution of the BH torus 
system. 
We refer the interested reader to  Ref.~\cite{Just:2014fka} 
for more details on the simulation setup. 

\begin{figure}[]
  \includegraphics[angle=0,width=0.95\columnwidth]{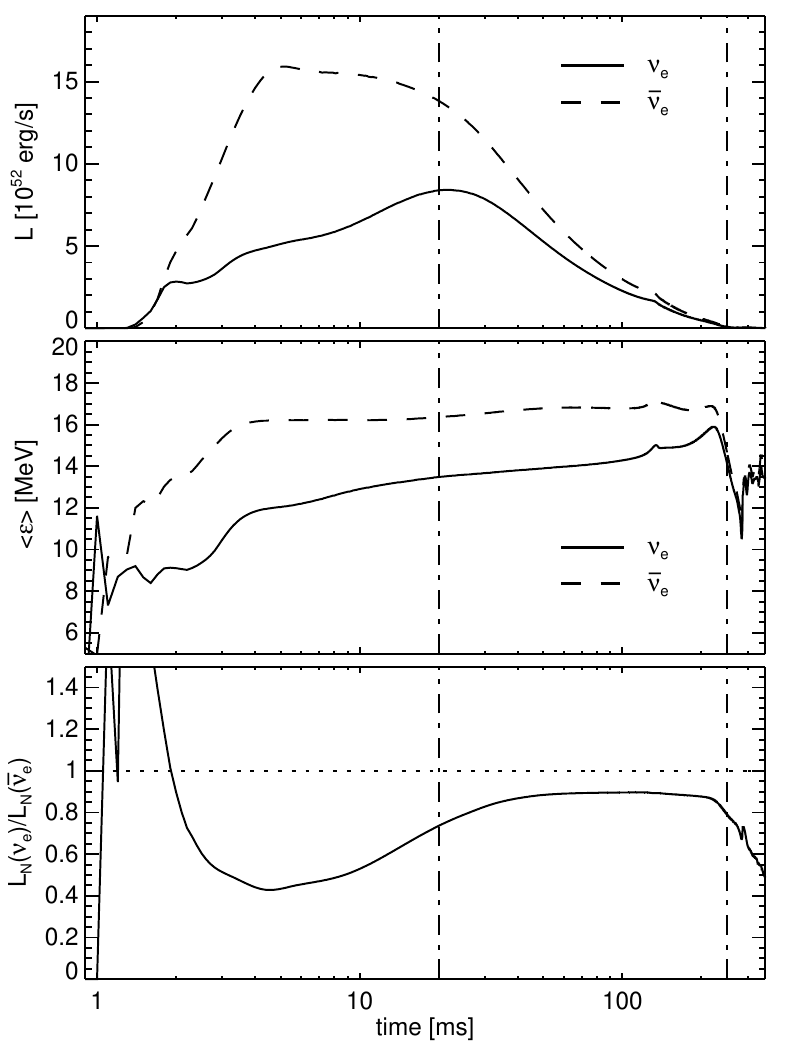}     
  \caption{
  Evolution of $\nu_e$ and $\bar\nu_e$ energy luminosities (top panel) and
  average energies (middle) as functions of time for the model M3A8m3a5. 
  The neutrino properties have been extracted at a radius of 500 km.
  The bottom panel shows the ratio of the number luminosities
  of $\nu_e$ and $\bar\nu_e$, which indicates that the
  torus on average protonizes (apart from the first $\sim 2\,$ms)
  until neutrino emission becomes inefficient. The vertical lines approximately mark the three different stages of the torus 
  evolution, see text for more details. 
  }
  \label{fig:v_global}
\end{figure}

Depending on the initial condition, an accretion torus evolves during time scales 
ranging from tens of milliseconds to seconds. It loses mass by accreting onto 
the BH and by thermally and viscously driven outflows. 
The evolution of such a massive and dense torus can be divided
mainly into three different stages, as shown in Fig.~\ref{fig:v_global}. 
Initially, the environment is dense enough to produce optically thick conditions 
for neutrinos, for which the latter are partially trapped and advected with the flow
and neutrino cooling is less efficient. This first phase lasts for about $\mathcal{O}(10)$~ms.

As the mass of the torus decreases and the density drops, the 
phase of a ``neutrino-dominated accretion flow'' begins. During this phase, neutrinos 
radiate away most of the gravitational energy that is converted into internal 
energy via viscous heating. 
As the mass, density and temperature of the torus 
continue to decrease, the neutrino production rate decreases until neutrino cooling 
becomes inefficient again at $t\approx 0.2-0.3$~s. 
This can be seen in the top panel of Fig.~\ref{fig:v_global} which shows the evolution
of the energy luminosities of both $\nu_e$ and $\bar\nu_e$ in the laboratory frame at a spherical radius of 500\,km.
The bottom panel of Fig.~\ref{fig:v_global}, displaying the ratio of the number luminosities
of $\nu_e$ and $\bar\nu_e$, shows that during the entire evolution phase of the torus
when neutrinos are efficiently produced, the torus on average continues to protonize (apart from the first $\sim 2\,$ms during
which the electron fraction in the densest parts of the torus settles from its initial value of $0.1$ to a new, slightly
lower weak equilibrium value).

The behavior of the average electron fraction of the torus and its temporal change (the latter being
given by the ratio of number luminosities, $L_{\mathrm{N},\nu_e}/L_{\mathrm{N},\bar\nu_e}$, see
bottom panel of Fig.~\ref{fig:v_global}) can be understood as follows: Accreting torus material at
all times tends to achieve $\beta$-equilibrium (e.g.~Ref.~\cite{2003ApJ...588..931B}). The electron fraction
corresponding to this equilibrium (and therefore the actual $Y_e$ of the torus) remains rather low
($\ll 0.5$) during the first two evolutionary phases as a result of self-regulation between
viscous heating and neutrino cooling into a state with semi-degenerate 
electrons~\cite{Chen:2006rra,Just:2014fka,Siegel:2017nub}. However, as the torus becomes more and more
diluted due to accretion onto the BH, the electron degeneracy is (on average) lifted, causing
the gas to favor a more symmetric $\beta$-equilibrium regarding the abundance of neutrons and
protons, i.e.~higher electron fractions.

\begin{figure*}[]
  \includegraphics[angle=0,width=1.95\columnwidth]{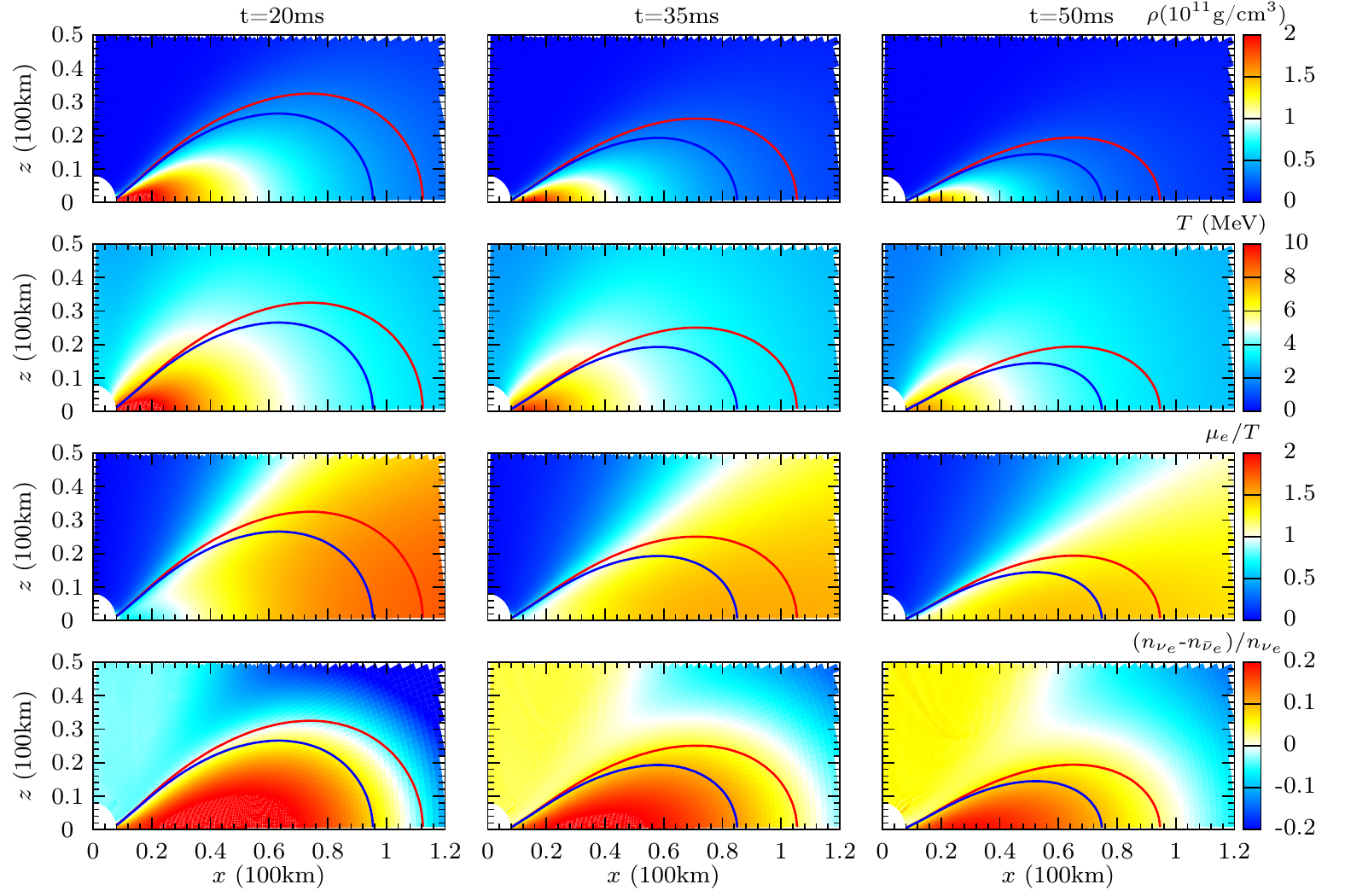}     
  \caption{BH-torus remnant properties for  model M3A8m3a5 around the $z$-axis at 20, 35 and 50 ms  (from left to right)
  as functions of $x$ and $z$ (assuming cylindrical symmetry around the $z$-axis).    
  First row: Baryon mass density $\rho$.  
  Second row: Temperature $T$. 
  Third row: Degeneracy parameter $\mu_e/T$ of electrons with
  $\mu_e$ being the electron chemical potential.
  Fourth row: Relative electron neutrino lepton number 
  ($n_{\nu_e}-n_{\bar{\nu}_e})/n_{\nu_e}$. 
  The degeneracy parameter in the innermost torus region slightly increases over time
  as the torus evolves (see text for details).
  The BH-torus evolves from a configuration where $n_{\bar{\nu}_e} > n_{\nu_e}$ to a configuration where 
  $n_{\bar{\nu}_e} < n_{\nu_e}$ around the polar axis. 
  Also shown are the emitting surfaces of $n_{\nu_e}$ and 
  $n_{\bar{\nu}_e}$ which are computed as described in Sec.~\ref{sec:nusurf} 
  and are marked in red and blue, respectively.  
  \label{fig:fnu_dens}}
\end{figure*}

The particularly low values of $L_{\mathrm{N},\nu_e}/L_{\mathrm{N},\bar\nu_e}$ during the early,
optically thick phase (indicating strong protonization) reflect the circumstance that torus material
is accreted before it is able to reach $\beta$-equilibrium, as a result of the diffusion time scale
being longer than the viscous accretion time scale. Afterwards, once the optical depth of the
torus sufficiently drops, $L_{\mathrm{N},\nu_e}/L_{\mathrm{N},\bar\nu_e}$ reaches values just below unity,
which means that the protonization takes place more slowly.

After the first two evolutionary stages, the system transits to the so-called ``advection
dominated accretion flow.''  During this phase, viscous heating leads to large-scale convective
motions and pushes the torus towards expansion.

Related to the mass ejection during the torus evolution, 
in the first two phases, the high-neutrino luminosities emitted by the inner 
parts of the torus heat and irradiate the outer and less dense layers of the torus. 
By doing so they form an outflow similar to the neutrino-driven wind 
present in CCSNe (see e.g., Refs.~\cite{Qian:1996xt,Arcones:2006uq}). 
The mass loss rate, the thermodynamic conditions and the neutron-to-proton 
ratio strongly depend on the neutrino emission properties of the torus. 

In the last stage, once neutrino cooling becomes inefficient, 
viscous angular momentum transport, together with 
viscous heating and nuclear recombination drive an inflation 
of the torus. This also leads to mass outflows which, however, 
are more massive near the equatorial plane.
Such a viscously driven wind proceeds gradually and 
the matter becomes gravitationally unbound at large radii, with velocities lower 
than that of the supersonic neutrino-driven wind. More matter can be ejected during this
phase. 

For model M3A8m3a5  studied in this paper, 
the total mass of the neutrino-driven ejecta is approximately 
$1.47 \times 10^{-3}\ M_\odot$ compared to the total outflow mass 
which is about $66.2 \times 10^{-3}\ M_\odot$~\cite{Just:2014fka}.
Since we are interested in the effect of neutrino flavor conversions
on the ejecta properties, throughout the paper, 
we will focus on the early outflow driven by 
neutrinos (i.e., $t \lesssim 60$~ms).

\subsection{Electron neutrino lepton number and other remnant emission properties}
\begin{figure*}[]
  \includegraphics[angle=0,width=2.05\columnwidth]{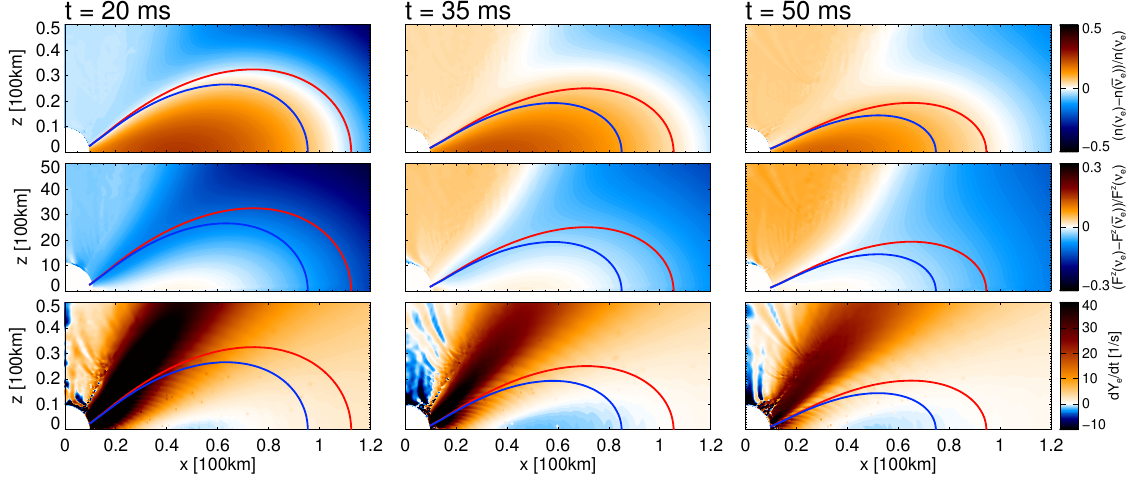}     
  \caption{BH-torus remnant properties for model M3A8m3a5 at 20, 35 and 50 ms (from left to right)
  as functions of $x$ and $z$ (assuming cylindrical symmetry around the $z$-axis).    
  First row: Relative ELN, ($n_{\nu_e}-n_{\bar{\nu}_e})/n_{\nu_e}$.  
  Second row: Relative ELN flux density $(F^z_{\nu_e}-F^z_{\bar{\nu}_e})/F^z_{\nu_e}$ 
  along the $z$ direction [see Eq.~\eqref{eq-nflux} for the definition]. 
  Third row: Local protonization rate $dY_e/dt$.
  The reduced protonization rate as the torus evolves and the different 
  emission geometry of the $n_{\nu_e}$ and $n_{\bar{\nu}_e}$
  surfaces (red and blue curves) lead to the growing excess of $\nu_e$ compared to $\bar\nu_e$
  around the $z$-axis.} 
  \label{fig:torusev}
\end{figure*}

The bottom panel of Fig.~\ref{fig:v_global} shows that the relative rate of protonization
changes as a function of time for $t\lesssim 60$~ms (i.e., the time window relevant for
the neutrino-driven ejecta). 
We, therefore, show in Fig.~\ref{fig:fnu_dens} 
the matter density $\rho$, the temperature $T$, the
degeneracy parameter $\mu_e/T$ with $\mu_e$ being the electron chemical potential,
and the ELN  ($\equiv n_{\nu_e}-n_{\bar\nu_e}$),
as functions of $x$ and $z$ (assuming cylindrical symmetry) from top to bottom.
Each quantity is shown for three selected snapshots at 
$t=20, 35$ and $50$~ms (from left to right) to illustrate the evolution of the torus conditions. 
The surfaces where $\nu_e$ and $\bar\nu_e$ decouple
are also shown in red and blue, respectively
(see Sec.~\ref{sec:nusurf} for more details).

As the torus continuously accretes onto the BH, 
both $T$ and $\rho$ decrease.
Consequently, the size of both neutrino surfaces shrinks.
However, the electron degeneracy $\mu_e/T$ in the innermost
part of the torus increases from $\mu_e/T<1$ to $\mu_e/T\sim 1$
as the torus evolves and neutrino cooling becomes more efficient
with decreasing optical depth. This increase of the electron 
degeneracy~\footnote{We note that a local increase of the electron 
degeneracy is not inconsistent with the previous statement that 
this quantity globally (i.e., averaged over the entire torus) decreases.}
leads to 
a relatively larger ratio of the electron capture rate to the positron capture rate.
Since most of the neutrinos ending up in the polar region are emitted from this inner region
of the torus, this has consequences on the ELN above 
the neutrino surfaces. The bottom panels of Fig.~\ref{fig:fnu_dens}
show that, at 20 ms, the whole region above the torus is characterized by 
$n_{\bar{\nu}_e} > n_{\nu_e}$. The torus gradually evolves 
towards a configuration where $n_{\nu_e}>n_{\bar{\nu}_e}$ in the polar region
at later times.
The main reason for having a $\nu_e$ excess in the polar region
is due to a geometrical effect.
As the $\nu_e$ surface with a conical shape is more extended than the 
$\bar\nu_e$ surface with nearly the same half-opening angle (Fig.~\ref{fig:fnu_dens}), 
more $\nu_e$'s are emitted towards 
the polar region than $\bar\nu_e$'s from their respective surfaces. 
This  results in a $\nu_e$ excess when
the torus is only slightly protonizing at later times.
Figure~\ref{fig:torusev} provides more insight into this evolutionary 
effect as a consequence of the neutrino transport conditions around the torus. 
As matter flows
towards the BH, it protonizes ($dY_e/dt > 0$) in all of 
the near-surface regions of the torus at all times, while 
the high-density inner parts have achieved a steady state
condition ($dY_e/dt \approx 0$) or neutronize with a very
low rate.

Nevertheless, it is only at early times that all the volume above
the neutrino surfaces is dominated by the number densities
and number fluxes of $\bar\nu_e$ (Fig.~\ref{fig:torusev}, left panels).
In contrast, at later times ($t \gtrsim 25$\,ms)
a growing conical volume around the rotation axis develops
an excess of $\nu_e$ in number density and number flux.
The reason is twofold: First, the decreasing rate of 
protonization with progressing evolution (compare left and
right columns of Fig.~\ref{fig:torusev}) near the torus surface reduces the
difference between the overall higher $\bar\nu_e$ number
flux compared to the $\nu_e$ number flux, as well as locally at the 
neutrino surfaces. Second, the 
different emission geometry of the $\nu_e$ and $\bar\nu_e$
surfaces plays an increasingly more important role:
because the neutrino surface of $\nu_e$ is more extended,
it irradiates the region around the rotation axis from a
wider angle than the $\bar\nu_e$ surface does. Both effects 
combined lead to the growing excess of $\nu_e$ compared to
$\bar\nu_e$ around the $z$-axis.

In the BH-torus model M3A8m3a5,
the transition between the two regimes from polar $\bar{\nu}_e$ excess to 
polar $\nu_e$ excess happens at about 25 ms. 
As we will see in Sec.~\ref{sec:instability}, this has important consequences on
the flavor conversions of neutrinos. 
We note here that such a transition is a generic feature seen in 
all BH-torus models in Ref.~\citep{Just:2014fka}, while only the 
transition time depends on the model parameters.

\subsection{Neutrino emission surfaces}\label{sec:nusurf}
Since the inner torus is dense enough to trap neutrinos, we can define the
$\nu_e$ and $\bar\nu_e$ emitting surfaces to approximate the boundaries above which neutrinos
can be considered as free-streaming particles, similarly to the neutrinosphere 
usually defined in CCSNe. Noticeably, the concept of emitting surfaces 
is nothing more than a formal definition; it is, however, useful for the
flavor instability study done in this work. 
In CCSNe, the ``neutrinosphere'' is  defined as 
the surface where the optical depth along the radial direction is about
 2/3. However, in the torus  case the optical depth becomes 
direction-dependent since the
geometry is highly non-spherical. It is therefore difficult to unambiguously define a 
direction for calculating the neutrino emission surface.

We therefore adopt a simpler approach guided by CCSN simulations.
At any location, the neutrino number flux $\mathbf{F_{\nu_\alpha}}$ and number density $n_{\nu_\alpha}$
of the flavor $\nu_\alpha$ are given by ($\hbar=c=1$)
\begin{equation}\label{eq-nflux}
\mathbf{F_{\nu_\alpha}}=\frac{1}{(2\pi)^3}
\int d\Omega dE\ \mathbf{\hat p} E^2 f_{\nu_\alpha}(\mathbf{p}),
\end{equation}
and 
\begin{equation}\label{eq-nden}
n_{\nu_\alpha}=\frac{1}{(2\pi)^3}
\int d\Omega dE\ E^2 f_{\nu_\alpha}(\mathbf{p}),
\end{equation}
where $d\Omega$ is the differential solid angle, 
$E$ and $\mathbf{p}$ are the neutrino energy and momentum
and $f_{\nu_\alpha}(\mathbf{p})$ is the
neutrino phase-space  distribution function.

By examining the CCSN simulations from Ref.~\citep{Fischer:2009af}, 
we found that the location of neutrinospheres obtained 
by using the definition based on the optical depth  agrees well with the 
location where the flux factor $j_{\nu_\alpha}\equiv|\mathbf{F}_{\nu_\alpha}|/n_{\nu_\alpha} =1/3$.
In fact, when the neutrino 
distribution is isotropic and neutrinos are trapped, the flux factor is zero, 
whereas when neutrinos start to free stream, the flux factor increases 
until it approaches unity at large distance from the emitting 
surface. 

\begin{figure}[]
  \includegraphics[angle=0,width=0.95\columnwidth]{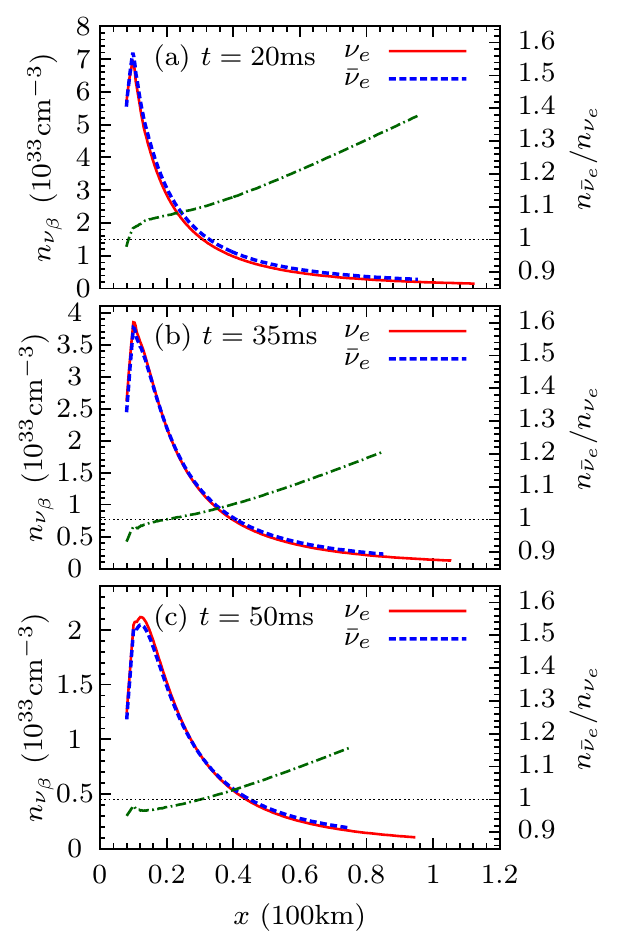}     
  \caption{Number density of $\nu_e$ and $\bar\nu_e$ on their 
  respective emitting surfaces at $t=20$~ms [panel(a)], 
  35~ms [panel(b)] and 50~ms [panel(c)].
  As the remnant evolves, the inner torus emits more $\nu_e$
  than $\bar\nu_e$ despite protonizing as a whole. 
  The ratio of $n_{\bar\nu_e}/n_{\nu_e}$ (green curves) is also 
  plotted so that this transition is more clearly visible. The black dotted line denotes $n_{\bar\nu_e}/n_{\nu_e}=1$
  to guide the eye.}
  \label{fig:v_surf}
\end{figure}

Figure~\ref{fig:v_surf} shows the neutrino number density of
$\nu_e$ and $\bar\nu_e$ on their respective neutrino
surfaces at 20, 35, and 50~ms as functions
of $x$. 
The effect of the reduced protonization discussed
in the previous section is also visible here.
At $20$~ms, the local $\bar\nu_e$ number density is
higher than that of $\nu_e$ with the same $x$.
However, at later times ($t\gtrsim 35$~ms), the $\nu_e$ number density 
in the inner torus region exceeds the $\bar\nu_e$ one.

In the rest of this paper, we will take these 
$\nu_e$ and $\bar\nu_e$ surfaces as the inner boundaries
where neutrinos are emitted and propagate freely afterwards.
The impact of the above transition
on the flavor instability will be discussed in the next section.

\section{Neutrino flavor conversions in compact binary mergers}\label{sec:nuoscillations}
In this section, we introduce the dispersion relation (DR)
in the neutrino flavor space. 
Our results on the flavor instabilities regarding fast pairwise 
conversions are also presented. 

\subsection{Dispersion relation in the flavor space}
In the free-streaming regime,
the equations of motion (EoMs) governing the flavor evolution of neutrinos 
are usually expressed in
terms of the density matrix $\varrho$,
which encodes the flavor occupation numbers in the 
diagonal terms and the flavor correlations in the off-diagonal
terms. The EoMs are
\begin{equation}
(\partial_t+\mathbf{v}\cdot\partial_{\mathbf x})\varrho = -i[H,\varrho]\ ,
\end{equation}
where  $\mathbf{v}=(\sin\theta\cos\phi,\sin\theta\sin\phi,\cos\theta)$ 
is the velocity of an ultra-relativistic neutrino and its 4-vector is $v^\mu=(1,\mathbf{v})$. 
The Hamiltonian, $H$, consists of the vacuum term which takes into account the flavor mixing
of neutrinos in vacuum~\cite{Olive:2016xmw}, the matter term describing the 
coherent forward scattering of neutrinos with the matter background~\cite{Mikheyev:1986gs,Mikheev:1986if} 
and the $\nu$--$\nu$ term taking into account the interactions of neutrinos with their
own background~\cite{Fuller:1987,Sigl:1992fn,Pantaleone:1992eq}.

In this work, we are interested in investigating the role of fast flavor
conversions~\cite{Sawyer:2005jk,Sawyer:2008zs,Sawyer:2015dsa}. 
Therefore, we will neglect the vacuum term in the Hamiltonian as well as the 
dependence of the neutrino energy and rely on a two-flavor framework
$(\nu_e,\nu_x)$ where $\nu_x$ is a linear combination of the non-electron 
flavors~\cite{Izaguirre:2016gsx,Wu:2017qpc}.

Expressing the neutrino density matrix as a function of the ``flavor isospin'' 
$\xi$ and the occupation number $f_{\nu_\alpha}$ for the neutrino
flavor $\nu_\alpha$:
$\varrho=[(f_{\nu_e}+f_{\nu_x})+(f_{\nu_e}-f_{\nu_x})\xi]/2$ 
($\bar\varrho=-[(f_{\bar\nu_e}+f_{\bar\nu_x})+
(f_{\bar\nu_e}-f_{\bar\nu_x})\xi^*]/2$)
for neutrinos (antineutrinos) and
introducing  the metric $\eta^{\mu\nu}={\rm diag}(1,-1,-1,-1)$,
the Hamiltonian for $\xi(\mathbf{v})$ can be written as
\begin{equation}\label{eq-eomnfis}
H = v^\mu \lambda_\mu \frac{\sigma_3}{2}+ \int d\Omega^\prime 
v^\mu v^\prime_\mu \xi(\mathbf{v}^\prime)g(\mathbf{v}^\prime)\ .
\end{equation}
The term $v^\mu \lambda_\mu = \lambda_0 - \mathbf{v} \cdot {\bm\lambda}$, 
$\bm\lambda=\lambda_0\mathbf{v_f}$, 
where $\mathbf{v_f}$ is the local fluid velocity, and $\lambda_0 = \sqrt{2} G_F n_e$ where
$n_e$ is the net electron number density. In the following, we will work in the 
frame corotating with $\bm\lambda$ and therefore
take $v^\mu \lambda_\mu = \lambda_0$.

The neutrino angular distribution potential $g(\mathbf{v})$ per unit length per unit solid 
angle is proportional to the ELN angular distribution
\begin{equation}
\label{eq:disk}
g(\mathbf{v})\!=\! \sqrt{2}G_F
\left(\Phi_{\nu_e} - \Phi_{\bar{\nu}_e}\right)\ ,
\end{equation}
where $\Phi_{\nu_\alpha}=dn_{\nu_\alpha}(\mathbf{v})/d\Omega$.

We are now interested in looking for non-null off-diagonal terms in the density matrix which 
could eventually arise,  giving origin to fast conversions. 
Hence, we linearize the EoM~\cite{Banerjee:2011fj,Raffelt:2013rqa} and
follow the evolution of the off-diagonal term $S$ in $\xi$, neglecting terms 
larger than $\mathcal{O}(|S|)$:
\begin{equation}
\xi=\left(\begin{matrix}
1 & S\\
S^* & -1\\
\end{matrix}
\right).
\end{equation}

We make the ansatz that $S$ evolves as  
$S(\mathbf{v},t,{\mathbf x})=Q(\mathbf{v},\omega,\mathbf k)
e^{-i(\omega t- {\mathbf k}\cdot{\mathbf x})}$. The  EoM becomes~\cite{Izaguirre:2016gsx}
\begin{equation}\label{eq-eom2}
v_\mu s^\mu Q(\mathbf{v},\omega,\mathbf k)
+\int d\Omega^\prime
v_\mu v^{\prime\mu}g(\mathbf{v}^\prime)
Q(\mathbf{v^\prime},\omega,\mathbf k)=0\ .
\end{equation}
Here, we have introduced the 4-vector 
$s^\mu\equiv (\omega-\lambda_0-\epsilon_0,
\mathbf{k}-{\bm\epsilon})$,
$\epsilon_0\equiv \int d\Omega g(\mathbf{v})$ and 
$\bm\epsilon \equiv \int d\Omega g(\mathbf{v})\mathbf{v}$. 
From the structure of Eq.~\eqref{eq-eom2}, one sees 
that the eigenfunction 
satisfying this equation is given by 
$Q(\mathbf{v},\omega,\mathbf k) \propto (v_\mu a^\mu)/(v_\mu s^\mu)$,
with $a^\mu$ being the coefficient of the eigenfunction solution.
A non-trivial solution for the $(\omega,\mathbf k)$ mode exists for 
non-zero $a^\mu$, if 
\begin{equation}\label{eq-DR-general}
{\rm det}[\Pi_{\mu\nu}(\omega,\mathbf k)]=0\ ,
\end{equation}
with $\Pi_{\mu\nu}(\omega,\mathbf k)=\eta_{\mu\nu}+\int d\Omega\ v_\mu v_\nu 
g(\mathbf{v})/(v_\mu s^\mu)$. The above equation corresponds to defining a DR in the 
flavor space.

If the solutions of Eq.~\eqref{eq-DR-general} are real, 
any initial perturbations in the flavor space do not grow. However, if any complex solution 
in $(\omega,\mathbf{k})$ satisfies the DR equation, then an exponentially 
growing instability may occur and lead to flavor conversions. 

In what follows, we will look for temporal instabilities, mainly originating from crossings in the 
ELN angular distribution, i.e.~look for 
complex solutions of $\omega$ with a given $\mathbf{k}$ satisfying Eq.~\eqref{eq-DR-general}~\cite{Izaguirre:2016gsx}.
The growth rate of the flavor instability will be given by $|{\rm Im}(\omega)|$.
We neglect the occurrence of spatial instabilities 
(occurring for $|{\rm Im}(k_i)|\neq 0$). In fact, the authors of Ref.~\cite{Wu:2017qpc}
found that the latter should cover a much smaller spatial region than the temporal 
instabilities. Also, the authors of Ref.~\cite{Capozzi:2017gqd} recently 
concluded that non-zero $|{\rm Im}(k_i)|$ alone 
might not lead to an exponentially growing instability.

\subsection{Instabilities in the flavor space}\label{sec:instability}

The authors of Ref.~\cite{Wu:2017qpc}, by approximating the merger remnant with a 
two-disk model, found that crossings in the ELN are present everywhere 
above the neutrino emitting surface.
We now intend to verify whether such conclusions still hold within a realistic torus configuration
evolving in time.
To this purpose, we investigate the DR [Eq.~\eqref{eq-DR-general}]
assuming neutrinos are emitted from their respective decoupling surfaces
defined in Sec.~\ref{sec:nusurf}.

\begin{figure*}[t] 
  \includegraphics[angle=0,width=1.95\columnwidth]{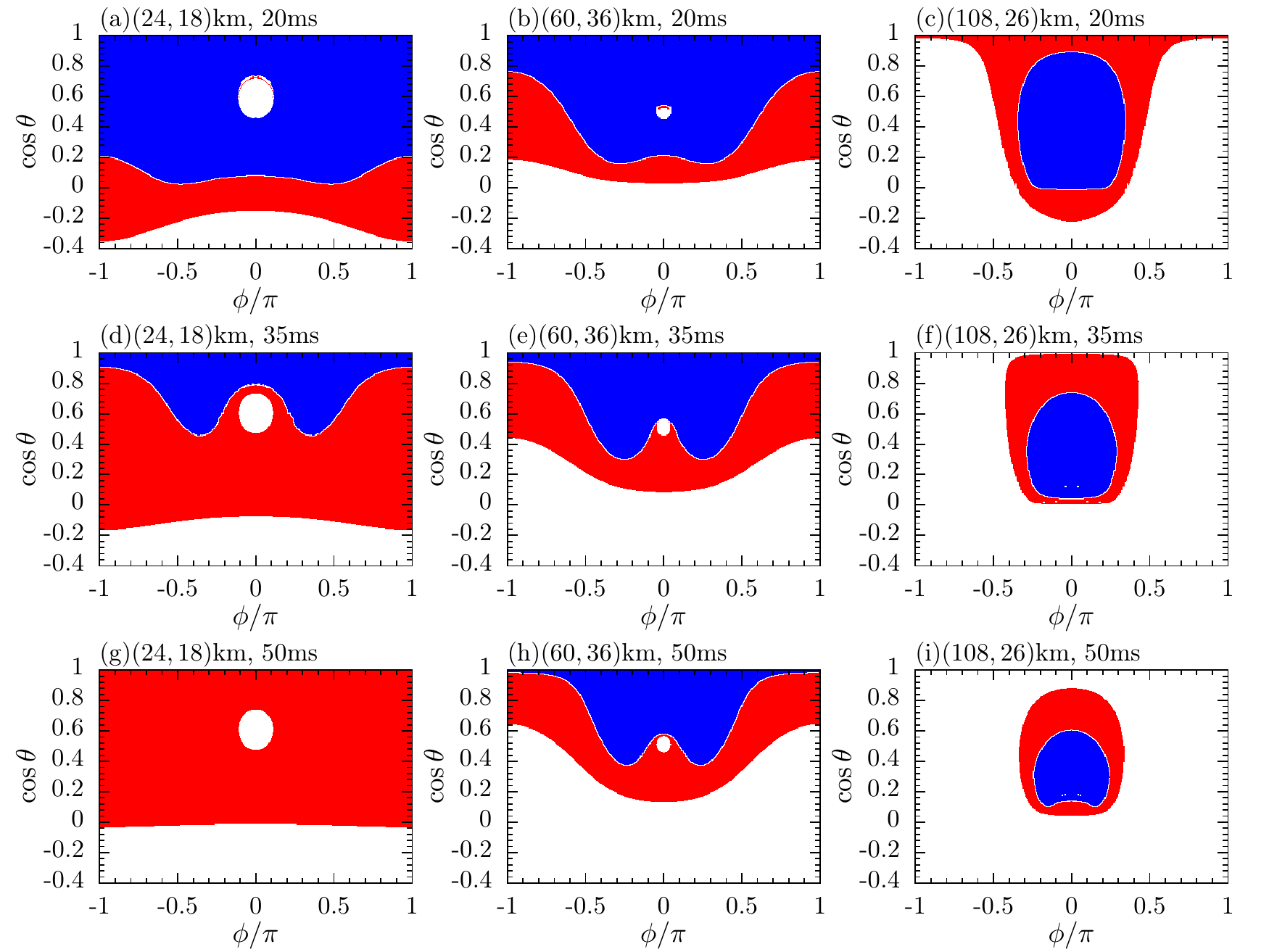}     
  \caption{Electron neutrino lepton number distribution as a function of 
  $\cos{\theta}$ and $\phi$ for different $(x,z)$ points above the $\nu_e$ 
  surface for the model M3A8m3a5 at 20~ms [panels (a)--(c)], 35~ms [panels (d)--(f)] 
  and 50~ms [panels (g)--(i)]. 
  In the blue (red) shaded areas, $\Phi_{\nu_e}-\Phi_{\bar{\nu}_e}<0$ 
  ($\Phi_{\nu_e}-\Phi_{\bar{\nu}_e}>0$). 
  The white regions mark the angular directions that do not cross the neutrino
  emitting surfaces and therefore the electron lepton number is zero. 
  \label{fig:ELN}}
\end{figure*}

Due to the approximate neutrino transport still adopted in merger simulations,
detailed neutrino angular distributions at
decoupling cannot be extracted. Hence, an assumption needs to be made
in order to obtain $g(\mathbf{v})$ above the
neutrino surfaces.

Our neutrino surfaces have been defined as the surfaces where the  flux factor
$j_{\nu_\alpha}=1/3$ (see Sec.~\ref{sec:nusurf}). One simple way to parametrize the
local neutrino angular distribution on the surface, that is consistent with this definition,
is to assume that the angular distribution grows linearly 
in $\cos\theta^\prime$, where 
$\theta^\prime$ is the angle with respect to the 
normal direction of the emission surface:

\begin{equation} 
\Phi_{\nu_e,\bar{\nu}_e}(\cos\theta^\prime)=\frac{n_{\nu_e,\bar\nu_e}}{4\pi}(1+\cos\theta^\prime)\ .
\end{equation}
One can easily verify that this angular distribution directly 
leads to $j_{\nu_e,\bar\nu_e}=1/3$ while respecting the
torus emission geometry.
At any location above
the $\nu_e$ surface, we can then calculate the corresponding 
neutrino angular distribution potential $g(\mathbf{v})$,
by applying the ray-tracing method (see e.g., Appendix~A in Ref.~\citep{Wu:2017qpc})
\footnote{We here
have neglected the neutrino ray bending effect due to general 
relativity. However, this effect should be minor in most of the regions, except those
immediately next to the BH.}.

Figure~\ref{fig:ELN} shows the resulting ELN distribution as a function of 
$\cos \theta$ and $\phi$ ($\theta$ is the angle relative to the $z$-direction, and $\phi$ is the angle relative to the 
$x$-direction) in  representative locations close 
to the inner [panels (a), (d), (g)], middle [panels (b), (e), (h)],
and outer [panels (c), (f), (i)] regions above the $\nu_e$ surface,
using the procedure described above for the model M3A8m3a5 
at 20, 35 and 50 ms.
The red and blue shaded areas distinguish between regions where the ELN potential is 
positive and negative, respectively.
The angular space where no neutrinos arrive from the emitting surfaces are
left in white.

One sees  from Fig.~\ref{fig:ELN} that, as 
the torus protonizes less, 
the stronger
$\nu_e$ emission from the inner torus leads to a smaller
solid angle  where $g(\mathbf{v})<0$ for the locations 
at the inner region. In particular, at later times, e.g.~$t=50$~ms, 
the ELN crossing in the inner region vanishes entirely [see panel (g)].
On the other hand, due to the persistently larger $\bar\nu_e$ 
emission in the outer torus, the ELN crossing still occurs 
for locations in the middle and outer parts above the $\nu_e$ surface.

We note here that, different from Ref.~\cite{Wu:2017qpc} where 
it was assumed that neutrino number densities were constant across their 
neutrino emitting surfaces, $n_{\nu_e}$ and $n_{\bar\nu_e}$
in this work are location dependent (see Fig.~\ref{fig:v_surf}).
Therefore, $g(\mathbf{v})$ is in general not uniform. The color
shading in Fig.~\ref{fig:ELN} is meant to only illustrate the structure of the ELN crossing.

One should also expect neutrinos to 
stream in the negative $\cos\theta$ direction  for locations above 
the $\nu_e$ emitting surface because of projection effects due to the  toroidal shape of the remnant. 
Moreover, a non-zero neutrino distribution in the negative $\cos\theta$ direction  
should also be expected due to neutrino scattering that results in the 
neutrino halo effect discussed in Refs.~\cite{Cherry:2012zw,Tamborra:2017ubu}.

\begin{figure}[t]
 \includegraphics[angle=0,width=1.\columnwidth]{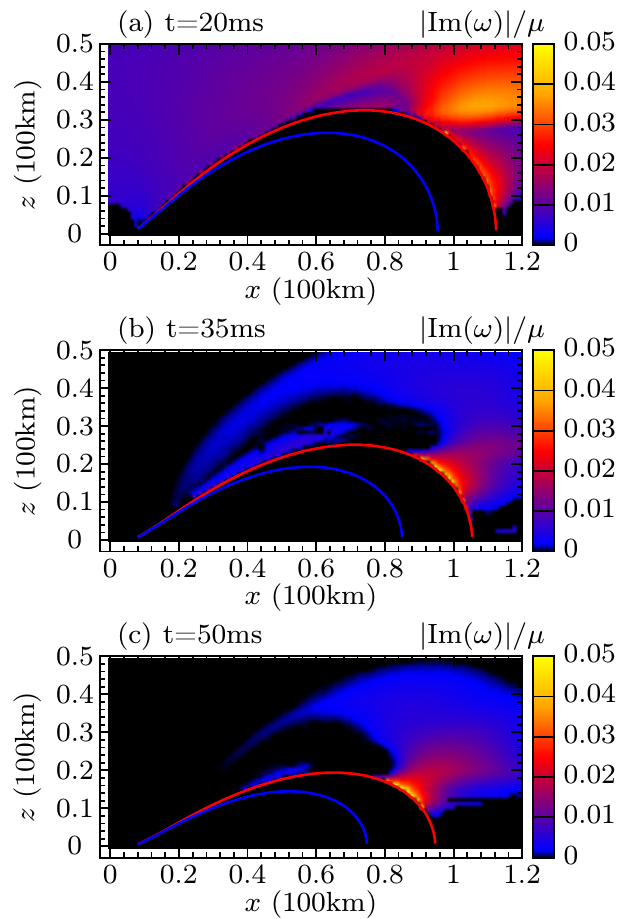}  
  \caption{Contour plot of $|\mathrm{Im}(\omega)|/\mu$ with $\mu = \sqrt{2} G_F n_{\nu_e}$ 
  in the $(x,z)$ plane above the $\nu_e$ surface for $k_x=k_y=k_z=0$ for $t=20, 35$ and $50$~ms
  from top to bottom respectively. 
  The $\nu_e$ ($\bar{\nu}_e$) neutrinosphere is marked in red (blue). At early times, fast conversions occur almost everywhere 
  above the $\nu_e$ surface. As the local $\nu_e$ abundance starts to be larger 
  than the $\bar{\nu}_e$ one, the regions above the torus where $|\mathrm{Im}(\omega)|/\mu \neq 0$ shrink considerably.
  The growth rates of the flavor instability 
  shown in all three panels range from 
  $10$~ns$^{-1}\lesssim |\mathrm{Im}(\omega)|\lesssim 1$~$\mu$s$^{-1}$.
  \label{fig:inst_plot}}
\end{figure}

Similarly to the conclusion in Refs.~\cite{Izaguirre:2016gsx,Wu:2017qpc},
complex solutions of the DR for a given $\mathbf{k}$ (i.e. the temporal flavor instabilities) 
exist whenever there is an ELN crossing. 
For the purpose of illustration, we show in Fig.~\ref{fig:inst_plot} the 
complex part of the DR solution [$\textrm{Im}(\omega)/\mu$] for the spatially homogeneous 
mode $\mathbf{k}=0$ for locations above the $\nu_e$ surface at 
$t=20$, 35, 50~ms, where $\mu=\sqrt{2}G_F n_{\nu_e}$.
The full solution of the DR for $\mathbf{k}=(0,0,k_z)$ at locations 
corresponding to the ELN distribution shown in Fig.~\ref{fig:ELN} is provided 
in the Appendix~\ref{sec:app-flavor}. 

Figure~\ref{fig:inst_plot} shows that  
fast conversions occur everywhere above the $\nu_e$ surface at $t=20$~ms. This can 
easily  be understood by looking at Fig.~\ref{fig:ELN} where for $t=20$~ms crossings in the ELN distribution
appear for any point above the torus when the torus is strongly protonizing.

At later times ($t>30$~ms), as the torus protonizes less and the local $\nu_e$ abundance 
starts to become larger and even dominates the $\bar{\nu}_e$ one around the polar region, we see
that the unstable region of $\mathbf{k}=0$ shrinks, particularly
in the region close to the pole and  immediately above the 
middle part of the $\nu_e$ surface.
At 50~ms, when the ELN crossing completely disappears in the
funnel region near the polar axis, the temporal instability is
  suppressed entirely. 
However, the local excess of $\bar{\nu}_e$ with respect to $\nu_e$
in the outer part of the disk still allows a 
large region for the flavor instability to exist for  $t=50$~ms.
We also note that $|\mathrm{Im}(\omega)|/\mu$ becomes smaller 
at later times. 
We note that the growth rates of flavor instability shown in 
Fig.~\ref{fig:inst_plot} range from 
$10$~ns$^{-1}\lesssim |\mathrm{Im}(\omega)|\lesssim 1$~$\mu$s$^{-1}$.
 
Our results confirm the findings of Ref.~\cite{Wu:2017qpc} and conclude 
that favorable conditions for fast flavor conversions  exist for 
the M3A8m3a5 torus. As discussed in 
Refs.~\cite{Sawyer:2005jk,Sawyer:2008zs,Sawyer:2015dsa}, the fact that
fast pairwise conversions could develop 
on time scales of ns to $\mu$s, much smaller than the
typical dynamical time scale of the system of $\gtrsim$~ms, 
means that neutrinos of 
different flavors could potentially reach flavor equilibration and 
share the same properties. 

As we will discuss in the next section, 
 nearly all the neutrino-driven trajectories 
are affected by neutrinos that cross the instability regions.
Hence, we will work under the assumption that flavor equilibration 
occurs because of pairwise conversions at $t\leq 50$~ms
to investigate the potential role of neutrino flavor conversions
in nucleosynthesis, instead of solving the full neutrino quantum kinetic equations in
the non-linear regime.  In fact new numerical 
tools need to be developed in order to incorporate fast pairwise conversions 
in the neutrino transport self-consistently. 
Our preliminary analysis here should serve as a test study to see 
whether more in-depth work on the modeling of pairwise neutrino conversions in 
binary neutron star merger remnants is needed.  

The stability analysis set forth in this section has been developed within a two neutrino 
flavor $(\nu_e,\nu_x)$ framework, as often adopted in the investigation of $\nu$-$\nu$ 
interactions; see e.g., references in Ref.~\cite{Mirizzi:2015eza}.
In the following, we will generalize our conclusions to a full three flavor 
framework. As discussed above, new numerical tools are needed to exactly estimate 
the expected flavor outcome; however it is conceivable that if a flavor instability 
develops within extremely small time scales this may 
lead to a full mixing of 
all three flavors.  As a consequence, in the following we will assume that 
flavor equilibration is reached for all three flavors
\begin{equation}
f_{\nu_e}=f_{\nu_\mu}=f_{\nu_\tau}=\frac{f_{\nu_e}^0}{3}\ ,
\label{eq:equil}
\end{equation}
for neutrinos that cross the unstable region shown in Fig.~\ref{fig:inst_plot}.
In the above equation~\eqref{eq:equil}, $f_{\nu_e}^0$ is the $\nu_e$ phase-space distribution function 
before flavor conversions occur.
An analogous relation is applied for antineutrinos.

\section{Impact of flavor equilibration on the nucleosynthesis in the neutrino-driven wind ejecta}\label{sec:nucleosynthesis}

\begin{figure*}[t]
  \includegraphics[angle=0,width=1.\columnwidth]{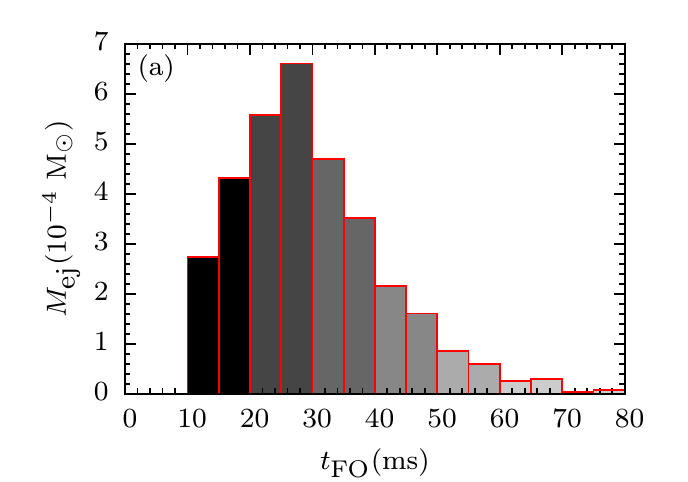}    
  \includegraphics[angle=0,width=1.\columnwidth]{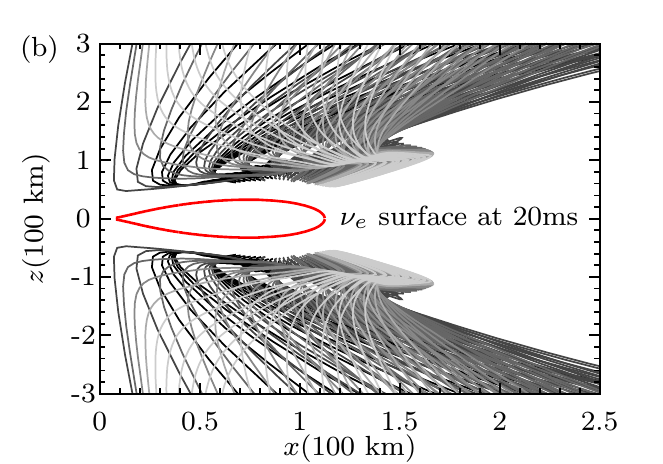}  
  \caption{Panel (a): Neutrino-driven mass ejecta as a function of the weak-interaction
  freeze-out time $t_{\rm FO}$ for the M3A8m3a5 model. 
  The freeze-out time is approximately 
  equal to the time where $Y_e$ reaches the asymptotic value $Y_{e,\rm asym}$
  before neutron capture starts. 
  The largest amount of neutrino-driven ejecta is emitted for $t< 35$~ms
  when the disk is strongly protonizing. 
  Panel (b): Trajectories of the neutrino-driven ejecta in the M3A8m3a5 
  model in the $(x,z)$ plane. 
  The $\nu_e$ emission surface at 20 ms is also plotted to guide the eye. 
  The color coding of the trajectories indicates the
  corresponding $t_{\rm FO}$ shown in panel (a).
  The trajectories ejected at early times originate closer to the polar region while 
  the later ejecta come from the outer 
  edges of the torus next to the equatorial plane. 
   \label{fig:massfraction}}
\end{figure*}

After briefly introducing the physics of heavy element nucleosynthesis in
the neutrino-driven wind in merger remnants, in this section
we explore the role of flavor equilibration on the nucleosynthesis outcome
of the neutrino-driven ejecta. 

\subsection{Neutrino driven wind in neutron star merger remnants}
Heavy elements in the neutrino-driven wind of the post-merger 
BH-torus remnant are produced in a way similar to the CCSN neutrino-driven wind:
matter recombines from free nucleons to form heavy nuclei as
the ejecta expand and cool. The detailed calculation of the nucleosynthesis process 
requires solving a large set of equations describing the
nuclear reaction network connecting different nuclei. 
In this work, we use the established nuclear reaction network suitable 
for the $r$-process nucleosynthesis calculation based on the
nuclear physics inputs of Ref.~\cite{Mendoza-Temis:2014mja}.
It contains 7360 nuclei and  all
relevant nuclear reactions.

Despite the complicated nuclear physics needed to model the quantitative nucleosynthesis results,
there are a few key quantities that determine the 
qualitative outcome (see e.g.,Refs.~\cite{Hoffman:1996aj,Lippuner:2015gwa}), namely the 
entropy, the ejecta expansion time scale and, most importantly, 
the electron abundance fraction per nucleon 
\begin{equation}
Y_e=\frac{N_e}{N_p+N_n}=X_p+\sum_{Z_A>2}\frac{Z_A}{A} X_A\ ,
\end{equation}
where $N_e$ ($N_p, N_n$) is the net electron (proton, neutron respectively) number density. 
$X_p$ and $X_A$ are the mass fractions of free protons and nuclei with charge numbers $Z_A\geq 2$.

In the early phases of the ejecta expansion when the temperature $T\gg 1$~MeV, 
matter mostly consists of free protons and neutrons, and 
the evolution of $Y_e$ is then set by the $\beta$-interactions of neutrinos with 
free neutrons and protons:
\begin{equation}
\nu_e+n \leftrightarrow p + e^-\ \mathrm{and}\ \bar{\nu}_e+p \leftrightarrow n + e^+\ .
\end{equation}
Therefore, the evolution of $Y_e$ during this phase can be approximated as
\begin{equation}
\frac{dY_e}{dt} \simeq (\lambda_{\nu_e}+\lambda_{e^+})Y_{n,f}-(\lambda_{\bar{\nu}_e}+\lambda_{e^-}) Y_{p,f}\ ,
\label{eq-yeevol}
\end{equation}
with $\lambda_i$ being the reaction rates and $Y_{n/p,f}\approx X_{n/p}$ the abundances of free nucleons. 

When the temperature drops to $T\gtrsim 1$~MeV before nucleons recombine
to $^4$He, the $e^\pm$ capture rates ($\lambda_{e^-,e^+}\propto T^{5}$) 
become much smaller than the neutrino absorption rates
($\lambda_{\nu_e,\bar\nu_e}$) and can later on be ignored.
Moreover, when both $\lambda_{\nu_e}$ and $\lambda_{\bar\nu_e}$ become smaller than the 
inverse of the radial expansion dynamical time scale of the ejecta, 
$\tau_{\rm dyn}^{-1}\simeq v_{{\rm ej},r}/r_{\rm ej}$,
where $v_{{\rm ej},r}$ and $r_{\rm ej}$
are the radial velocity and the radius for each given ejecta trajectory,
one can define this time as the weak-interaction freeze-out time, 
$t_{\rm FO}$.  At $t_{\rm FO}$, $Y_e$ 
of the trajectory roughly approaches an asymptotic value $Y^{\rm asym}_e$, 
until much later when the beta-decay of $r$-process nuclei sets in 
to further raise $Y_e$ (see Fig.~\ref{fig:trajectories} for a few examples).
Note that at $t_{\rm FO}$, all the neutrino driven ejecta
roughly expand along the radial directions.

Another quantity which is relevant for the subsequent discussion
is the so-called equilibrium electron fraction, $Y^{\rm eq}_e$, defined by
\begin{equation}\label{eq-yeeq}
Y_e^{\mathrm{eq}}=\frac{\lambda_{\nu_e}}{\lambda_{\nu_e}+\lambda_{\bar{\nu}_e}}\ .
\end{equation}
When the neutrino irradiation is very strong, 
$Y_e$ may reach $Y_e^{\mathrm{eq}}$ before the freeze-out.
Equation~\eqref{eq-yeeq} can be easily derived from Eq.~\eqref{eq-yeevol} by assuming $dY_e/dt=0$, 
neglecting $\lambda_{e^-,e^+}$, and taking $Y_{n,f}=1-Y_{p,f}=1-Y_e$~\cite{Hoffman:1996aj}.
In the typical CCSN neutrino-driven wind, this
condition generally holds as the ejecta overcome
the gravitational potential of the proto-neutron star  by 
neutrino energy deposition. However, in the next section, we will see
 that it is not generally true
for the neutrino-driven wind from post-merger BH-torus remnants,
as matter is more loosely bound in this case.

The amount of the neutrino-driven ejecta for the M3A8m3a5 model is shown as a function 
of $t_{\rm FO}$ in the left panel of Fig.~\ref{fig:massfraction}\footnote{For a comparison of 
the mass distribution histograms between the 
neutrino-driven ejecta and the viscously-driven ones, we refer 
the reader to Fig. 9 of Ref.~\cite{Just:2014fka}.}. Note that we have used for $t_{\rm FO}$ the 
same time coordinate as from the hydrodynamical simulation of model M3A8m3a5.
The trajectories for all the neutrino-driven ejecta in the $(x,z)$ plane 
are plotted in the right panel of Fig.~\ref{fig:massfraction}.
First, one sees that most of the neutrino-driven ejecta have the 
freeze-out time $t_{\rm FO} \simeq 15-40$~ms. Second, from the color coding, one can see that the trajectories 
ejected at early times originate mainly from regions next to the polar axis
while later ejecta come from the outer edges of the torus next to the equatorial plane. 

As the flavor instability exists
at any point above the $\nu_e$ surface at early times, and the outer 
part above the torus at later times, most
of the neutrino-driven ejecta with $t_{\rm FO}\lesssim 50$~ms
will be influenced 
by neutrinos that stream through the unstable regions
(see Fig.~\ref{fig:inst_plot} and Sec.~\ref{sec:instability} for comparison).
As a consequence, we will assume flavor equilibration [see Eq.~\eqref{eq:equil}]
happens for neutrino fluxes on ejecta trajectories at 
$t \leq 50$~ms in the following. 

\subsection{Impact of flavor equilibration on the element production}
We now explore the impact of flavor equilibration on the nucleosynthesis outcome
of the neutrino-driven ejecta.
Since the muon and tau (anti)neutrinos are produced only in the very innermost 
and dense regions of the torus, 
their luminosities are about 10 times lower 
than the ones of $\nu_e$ and $\bar{\nu}_e$~\cite{Just:2015dba,Janka:1999qu,Sekiguchi:2016bjd}.
We here neglect the non-electron flavors and perform nucleosynthesis calculations
by assuming that, when flavor equilibration occurs [see Eq.~\ref{eq:equil}], 
both the $\nu_e$ and $\bar\nu_e$ capture rates on nucleons are reduced to 
$1/3$ of their original values without oscillations. 

\begin{figure}[]
  \includegraphics[angle=0,width=1.\columnwidth]{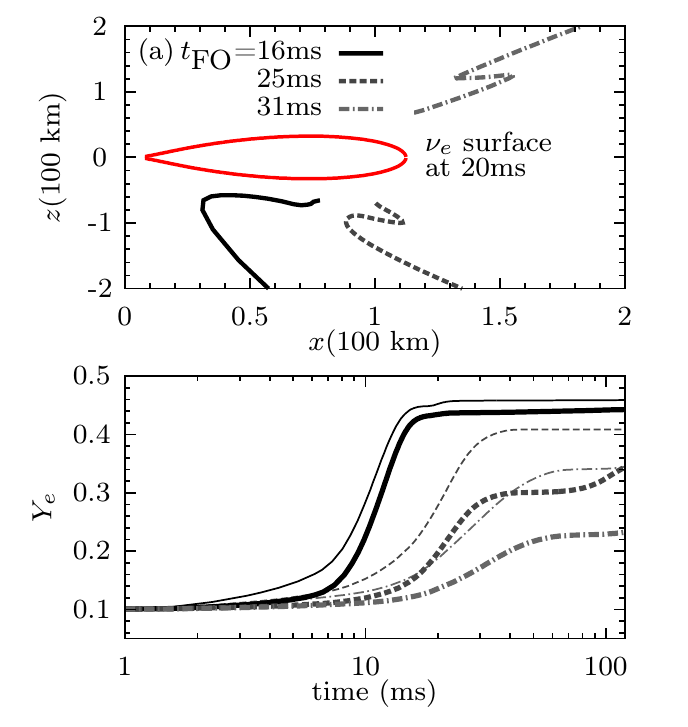}    
  \caption{
  Panel (a): Three selected neutrino-driven trajectories in the $(x,z)$ plane, labeled by 
  their respective $t_{\rm FO}$.
  The $\nu_e$ emission surface at 20 ms is also plotted to guide the eye. 
  Panel (b): Electron fraction $Y_e$ along the three selected trajectories 
  as a function of time; 
  $Y_e$ in the cases with (without) flavor equilibration is plotted with 
  the thick (thin) line. \label{fig:trajectories}}
\end{figure}

To discuss the impact of flavor equilibration on the electron fraction $Y_e$, 
we first examine three representative trajectories with $t_{\rm FO}=16$, 25 and 31~ms.
The top panel of Fig.~\ref{fig:trajectories} shows the selected trajectories in the $(x,z)$ plane.
The bottom panel of Fig.~\ref{fig:trajectories} shows  the evolution of $Y_e$ 
with and without flavor equilibration (thick and thin lines respectively).
The earlier ejecta with $t_{\rm FO}=16$~ms
originate from the region close to the pole and, therefore, are exposed to stronger
neutrino fluence. As a result, despite 
the strong reduction of the neutrino absorption rates,
$Y_e(t_{\rm FO})\approx Y_e^{\rm eq}$. This explains why a reduction of the neutrino 
rates due to flavor conversions has only little effect on the $Y_e$ evolution.
For the later ejecta, such as the ones with $t_{\rm FO}=25$ and $31$~ms, 
the asymptotic value of $Y_e$, $Y_e^{\rm asym}$, never 
reaches $Y_e^{\rm eq}$ even in the case without oscillations. 
Thus, the reduction of the neutrino capture rates due to flavor equilibration 
dramatically lowers the asymptotic value of $Y_e$ ($Y_e^{\rm asym}$) from $\sim 0.41$ and $0.34$ to
$\sim 0.3$ and $0.23$, respectively. Note that  $Y_e$ for the $t_{\rm FO}=25$~ms
trajectory shows a slight rise at $t \sim 80$~ms; this is due to the $\beta$-decays of 
neutron-rich nuclei during and after
the $r$-process.

\begin{figure}[]
  \includegraphics[angle=0,width=1.\columnwidth]{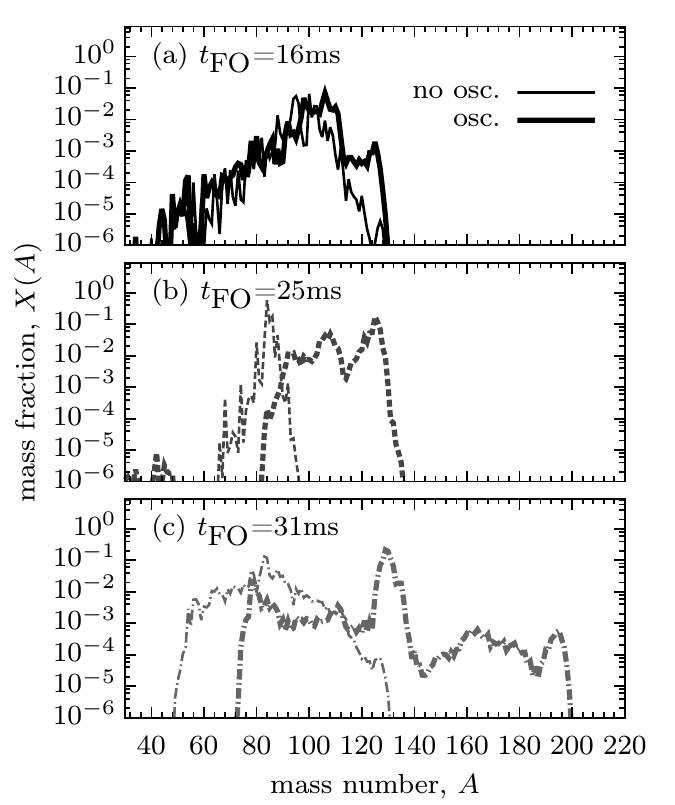}    
  \caption{Final nucleosynthesis outcome shown by mass fraction as a 
  function of the nuclear mass number for the same three selected trajectories
  shown in Fig.~\ref{fig:trajectories}.
  The cases with (without) flavor equilibration are plotted with the thick (thin) 
  lines. Flavor equilibration results in the production of elements with larger $A$. \label{fig:trajxA}}
\end{figure}

Figure~\ref{fig:trajxA} shows the final nucleosynthesis outcome
of these three trajectories. The mass fraction
$X(A)$ is plotted as a function of the nuclear mass number $A$.
As a consequence of the $Y_e$ evolution shown
in Fig.~\ref{fig:trajectories}, 
 there is only a small change in the nucleosynthesis pattern of the
earliest trajectory with $t_{\rm FO}=16$~ms, while 
the produced heavy nuclei in the later ejecta are shifted from peaking
around $A\sim 80$ to $A\gtrsim 130$, 
and even reaching the third peak $A\sim 195$ for the case with $t_{\rm FO}=31$~ms.

\begin{figure}[t]
  \includegraphics[angle=0,width=1.\columnwidth]{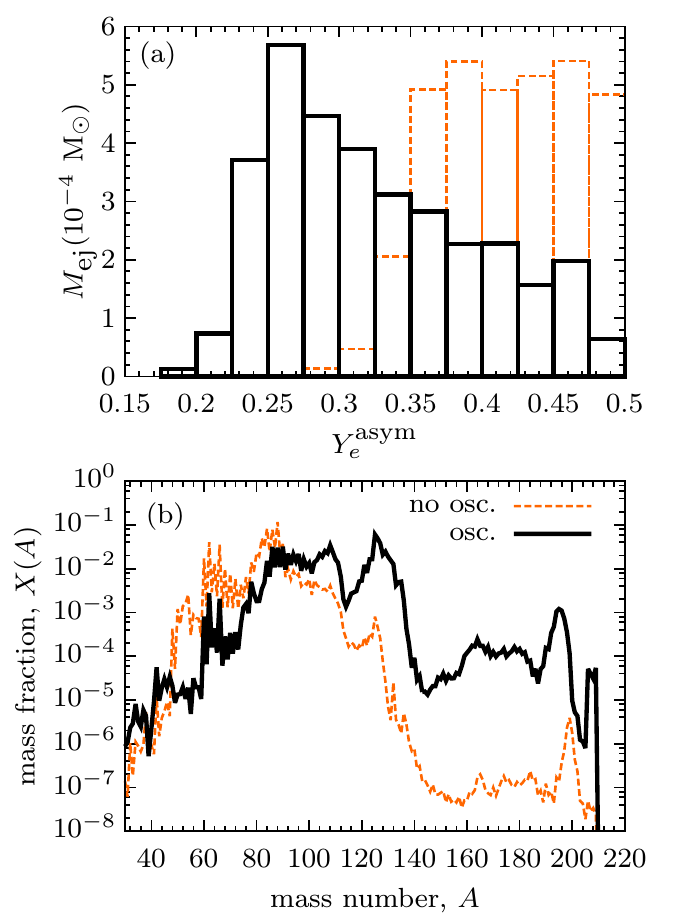}    
  \caption{Top panel: $Y_e^{\rm asym}$ distribution  
   with (black thick-solid) and without (orange thin-dashed) flavor equilibration. 
   Bottom panel: The corresponding mass fraction $X(A)$ as a 
   function of mass number $A$ for the whole neutrino-driven ejecta.
   Because of flavor conversions, the element production 
   shifts towards elements with heavier mass number.\label{fig:yenuc}}
\end{figure}

Figure~\ref{fig:yenuc} shows the ejecta masses as a function of the asymptotic $Y_e^{\rm asym}$ 
as well as the mass fraction for all the neutrino-driven trajectories shown in the 
right panel of Fig.~\ref{fig:massfraction} for the cases with and without 
flavor equilibration. 
As evident from the top panel  of Fig.~\ref{fig:yenuc}, the eventual occurrence of 
flavor equipartition greatly changes the $Y_e^{\rm asym}$ distribution of the ejecta,
from uniformly distributed in the range $Y_e \in [0.35,0.5]$ to being peaked 
around $Y_e \sim 0.25$ with a tail
distribution reaching $\sim 0.5$. 

The overall production of heavy elements
is therefore shifted from abundance peaks around $A\sim 80$ to $A\sim 130$, as shown in the bottom panel of Fig.~\ref{fig:yenuc}.
In addition, the production of nuclei above $A\sim 130$ is enhanced by more than 
a factor of 1000.\footnote{Note that for the no-oscillation case, 
the production of nuclei is slightly different 
with respect to Fig.~13 of Ref.~\cite{Just:2014fka}. 
This is due to the fact that we ignore ejecta
with $t_{\rm FO}<10$~ms from the torus in this work, 
as the torus is still going through an artificially high 
$\nu_e$ emission for $t_{\rm FO}<10$~ms~\cite{Just:2014fka}. 
By including the first 10 ms of the neutrino-driven ejecta, we can indeed reproduce the results in Ref.~\cite{Just:2014fka}, 
except for small differences due to different nuclear physics inputs.}

Our explorative study suggests that fast pairwise conversions may indeed
greatly affect heavy element production in the neutrino-driven wind of 
the merger remnant and strongly justifies further work in this direction. 
In particular, the enhancement of the production of lanthanides and the third-peak
nuclei can be substantial. 
This can potentially lead to interesting observational consequences
on the kilonova (macronova) ligthcurve, if the neutrino-driven wind dominates
the polar ejecta. For example, observations of the kilonova associated to the 
GW170817 event suggest blue (high $Y_e$) ejecta in the polar direction. Our results may support 
the interpretation that this observation points to a massive NS remnant that was stable for some time before 
collapsing to BH with some delay~\cite{Shibata:2017xdx}. In fact, the specific spectrum of the electromagnetic
signal may sensitively depend on the fraction of lanthanides~\cite{Barnes:2013wka,Tanaka:2013ana}.
If this should be the case, an increasing number of 
face-on observations of the kilonova ligthcurves along with theoretical
improvements in the modeling of binary mergers may also be able to put indirect constraints on 
fast flavor conversions and neutrinos.

We here assumed that flavor 
equilibration occurs for any time  $t \leq 50$~ms.
Given the change of sign of the ELN, our assumption may seem  extreme as the 
regions above the torus where the system is unstable shrink for $t>30$~ms (see Fig.~\ref{fig:inst_plot}). 
However, we stress again that  
the neutrino-dominated trajectories
although ejected at different times are always influenced
by the neutrinos crossing the instability regions. 
On the other hand, flavor equilibration will also reduce the heating and the amount of neutrino-driven 
ejecta. Matter can take longer to be ejected and  the real asymptotic $Y_e^{\rm asym}$ would likely be  sitting in
between our results with and without oscillations.
Given the potential major implications of neutrino conversions on heavy element synthesis, 
further exploration beyond the scope of this work is definitely needed
to fully understand the role of neutrino flavor conversions in post-merger nucleosynthesis.

\section{Conclusions and outlook}\label{sec:conclusions}

Binary neutron star mergers are neutrino-dense environments and likely sites 
for the rapid-neutron capture process. 
By adopting inputs from the hydrodynamical simulation of a binary neutron star merger 
with a black hole of $3 M_\odot$, dimensionless spin 
parameter $0.8$ and an accretion torus with mass
of $0.3 M_\odot$ (M3A8m3a5)~\cite{Just:2014fka}, 
for the first time, we have studied the neutrino 
emission properties as a function of time, investigated the conditions under which  fast pairwise 
conversions should develop in this environment, and examined the impact of flavor equilibration 
on the nucleosynthesis of heavy elements in the neutrino-driven ejecta. 

During the first $50$~ms of the neutrino-dominated accretion, when
neutrinos are efficiently emitted, the 
torus strongly protonizes initially. Then, it
gradually approaches a self-regulated semi-degenerate state
with a lower protonization rate.
Together with the geometry of the torus, this changes the 
electron neutrino lepton number distribution above the neutrino emitting surfaces 
 from being negative ($n_{\nu_e}<n_{\bar\nu_e}$) everywhere 
to exhibiting a $\nu_e$ excess in the funnel region around the polar axis
after $\sim 30$~ms. 
The fact that $\nu_e$ dominate with respect to $\bar{\nu}_e$  for $t> 30$~ms, implies the 
need for a time-dependent analysis of the flavor conversion phenomenology, which has never been addressed before. 

By performing a flavor stability analysis for different time snapshots 
of the remnant, we found that favorable conditions for fast conversions exist 
for every point above the $\nu_e$-emitting surface for $t<30$~ms. 
As $n_{\nu_e}$ starts to become larger than $n_{\bar{\nu}_e}$ around the polar axis, 
the region where the flavor instability can develop shrinks, 
 but it persists in the outer part above the torus where
crossings in the electron neutrino lepton number distributions occur.

Under the assumption that fast pairwise conversions 
lead to full flavor equilibration, we further investigated the impact of the  reduced 
neutrino absorption rates on nucleosynthesis in the neutrino-driven ejecta.
The $Y_e$ of the neutrino-driven
outflow can be largely reduced. 
Consequently, the production of nuclei with mass numbers larger than 130 can be
enhanced by more than a factor of 1000 with respect to the case where flavor conversions
are neglected. 
The enhanced production of lanthanides may also greatly
change the opacity of the neutrino-driven ejecta and
thereby affect the resulting kilonova ligthcurves.

In conclusion, our findings hint towards a relevant role of neutrino flavor conversions in 
binary neutron star merger remnants. The details of our findings should be taken 
with caution as our work was only meant to be exploratory and future work needs to be 
done to address the caveats adopted in this study.

One of the largest caveats is that we did not compute the exact flavor distribution due to 
fast pairwise conversions, but we assumed that flavor equilibration is reached given 
the temporal and spatial 
scale on which flavor instabilities are supposed to develop. 
Further numerical tools tackling fast pairwise conversions 
in a highly asymmetric environment need to be developed to this purpose.  
Furthermore, if fast conversions happen so close to the neutrino decoupling 
surfaces, a self-consistent feedback effect 
of flavor conversions on the ejecta composition and the merger dynamics needs to 
be carefully implemented.
The role of residual non-forward scatterings between neutrinos and matter above the
torus also needs to be examined.

We here only analyzed in detail the case of a remnant with
a central black hole. 
However, we note that different BH-torus configurations
in Ref.~\cite{Just:2014fka} all share the same qualitative behavior
of protonization and neutrino emission. Therefore, fast flavor conversions 
should occur in any BH-torus remnants and have similar impact
on the nucleosynthesis of heavy elements in the associated neutrino-driven wind. 
Remnants with a massive neutron star in the center 
should also be studied in this respect as the amount of neutrino-driven ejecta can 
be much larger in this case~\cite{Metzger:2014ila,Shibata:2017xdx}. 
We expect that the condition for fast flavor conversions to occur above the
remnant NS--disk system should also exist, based on the toy models studied in Ref.~\citep{Wu:2017qpc}.
The impact of flavor equilibration in such systems needs to be
carefully examined because substantial amount of non-electron flavor neutrinos
can be emitted from the central massive neutron star. 
Moreover, the condition of neutrino flavor conversions
during the dynamical merger phase should also be considered in the future as
several recent studies have shown that neutrinos can play an important role in 
driving the polar ejecta in the case of binary neutron star mergers~\cite{Wanajo:2014wha,Foucart:2016rxm}.
Together with the theoretical improvements, future gravitational-wave follow-up kilonova observations like
the recent detection of GW170817 will offer unique opportunities to
shed light on the role of neutrinos and their flavor conversions in compact binary mergers.

\section*{Acknowledgments}
M.-R.~W. and I.~T. acknowledge support from the Knud H{\o}jgaard Foundation, the Villum Foundation (Project No. 13164), and the
Danish National Research Foundation (DNRF91). The work of H.-T.~J. and O.~J. is supported by the European Research 
Council through grant ERC AdG 341157-COCO2CASA and the Cluster of Excellence ``Universe'' EXC~153. I.~T. 
and H.-T.~J. acknowledge support from the 
Deutsche Forschungsgemeinschaft through 
Sonderforschungsbereich SFB~1258 ``Neutrinos and Dark Matter
in Astro- and Particle Physics (NDM).
I.~T. and H.-T.~J. are also grateful to the Mainz Institute for 
Theoretical Physics (MITP) for its hospitality and support
during the completion of this work.  

\appendix


\section{Fast Pairwise Conversions: Instability regions}
\label{sec:app-flavor}
\begin{figure*}[t]
  \includegraphics[angle=0,width=1.8\columnwidth]{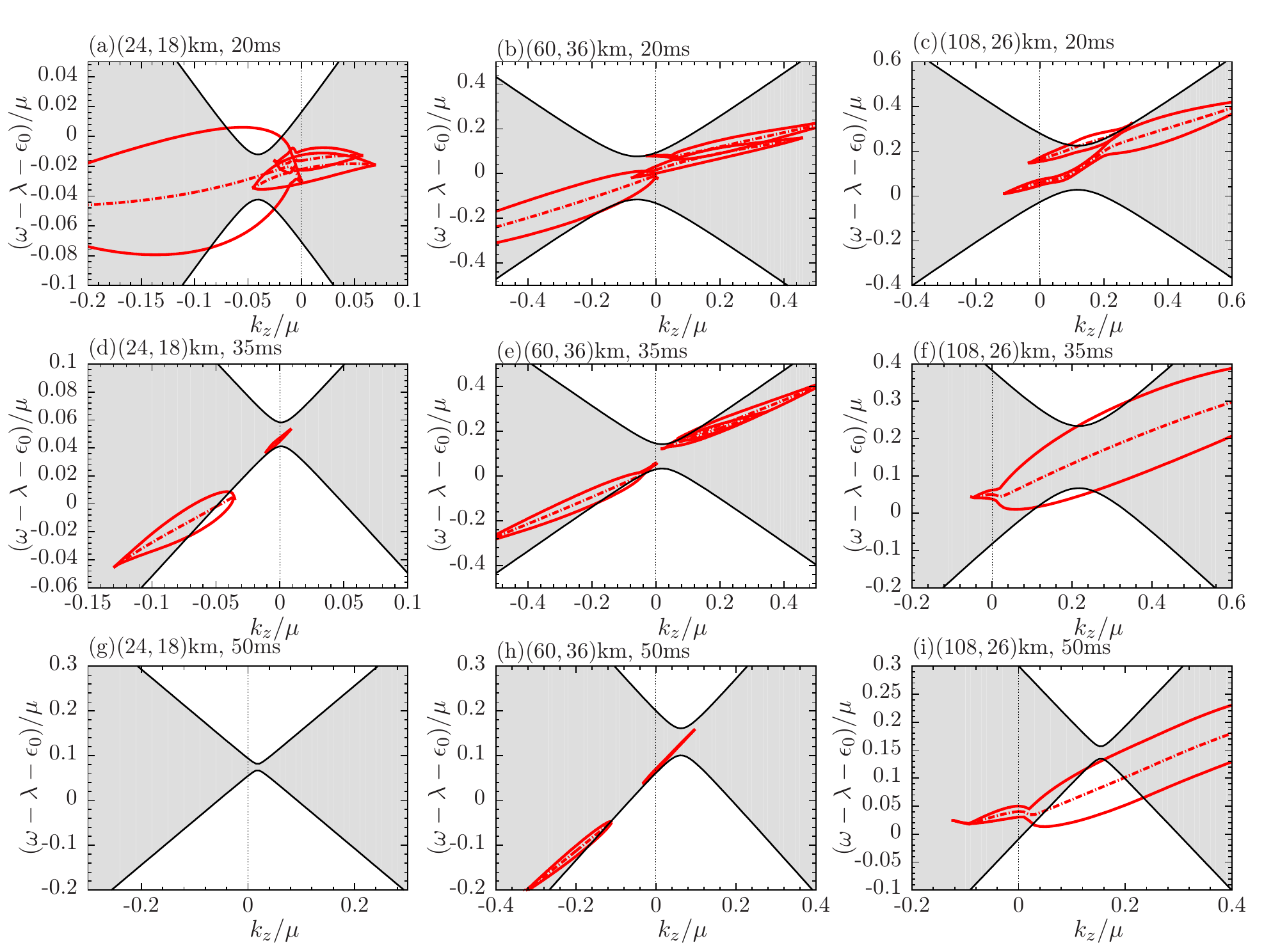}
  \caption{Dispersion relation of $\mathbf{k}=(0,0,k_z)$ for the ELN distribution 
  of $t=20, 50$~ms of the model M3A8m3a5. 
  For the complex $\omega$ solutions that lead to flavor instability, 
  $\mathrm{Re}(\omega)$ are plotted in 
  red dash-dotted curves and $\mathrm{Re}(\omega) \pm \mathrm{Im}(\omega)$ 
  are plotted with red solid curves. The  gray region 
  indicates the zone of avoidance for real $(\omega,k_z)$. 
  The system is unstable to fast conversions  for a large range of $k_z/\mu_0$ and it 
  is therefore unavoidable that fast pairwise conversions would occur. 
   \label{fig:complex_plot}}
\end{figure*}

In this appendix, we provide more details on the solution of the dispersion relation 
(Eq.~\eqref{eq-DR-general})at various locations above the merger remnant at different times.
Figure~\ref{fig:complex_plot} shows the solutions for 
$\mathbf{k}=(0,0,k_z)$ for the model M3A8m3a5 at the same locations and times
for given ELN distributions in Fig.~\ref{fig:ELN}.
The dash-dotted lines show the real part of the solutions while the continuous
lines represent the imaginary part. One can see that at 20~ms, unstable 
solutions always exist for a wide range of $k_z/\mu$
and shift from mainly at $k_z/\mu\lesssim 0$ to 
$k_z/\mu\gtrsim 0$ as one moves from the inner part above the remnant
to the outer part [panels (a)--(c)]. This is similar to what was found in Ref.~\cite{Wu:2017qpc}. 
At later times, $t = 35$ and $50$~ms, instabilities in general still exist [see panels (d)--(i)], 
however, the system becomes less unstable, particularly in regions close
to the polar axis as the ELN angular distribution become dominated by neutrinos (see Fig.~\ref{fig:ELN});
see e.g., panels (d), (g), and (h).
This is related to the fact that the the neutrino local density 
starts to exceed the antineutrino one and the BH-torus remnant approaches the
self-regulated equilibrium discussed in Sec.~\ref{sec:nuproperties}.




\end{document}